\documentclass[aps,prl,twocolumn,groupedaddress]{revtex4}  
\usepackage{graphicx}  
\usepackage{dcolumn}   
\usepackage{bm}        
\usepackage{amssymb}   
\usepackage{float}
\usepackage{color}

\begin{document}

\title{How do spins interact coherently with ultrashort laser pulses ?}

\widetext

\author{H\'el\`ene Vonesch}
\author{Jean-Yves Bigot}

\email{bigot@ipcms.u-strasbg.fr}

\affiliation{Institut de Physique et Chimie des Mat$\acute{e}$riaux de Strasbourg, UMR 7504 CNRS, Universit$\acute{e}$ de Strasbourg BP. 43, 23
rue du Loess, 67034 Strasbourg Cedex, France}


\begin{abstract}
We investigate the interaction of femtosecond laser pulses with spins including relativistic corrections. The
time-ordered magneto-optical signals corresponding to a pump-probe configuration  are calculated in the case of one electron submitted to a
magnetic field and evolving in eight levels of the fine structure of a hydrogen-like atom. Our simulations explain the origin of the
coherent magneto-optical response and ultrafast spin dynamics in ferromagnets excited by intense laser pulses as recently reported in Ni and
CoPt$_{3}$ ferromagnetic thin films. Our detailed analysis allows identifying the respective roles of the coherent spin-photon interaction and spin dynamics unraveling recent controversies about the laser induced ultrafast magnetization dynamics.
\end{abstract}

\maketitle


Ferromagnetic thin films can be modified on a femtosecond time-scale by ultrashort laser pulses \cite{Beaurepaire1996,Hohlfeld1997,Aeschlimann1997,Scholl1997,Kirilyuk2010}. This observed phenomenon, as well as the non-thermal optical control of magnetic order with light pulses \cite{Kimel2005}, is important as it promises major applications in the fields of data-storage and time resolved magnetic imaging for example \cite{Laraoui2007}. Understanding the underlying physical mechanisms requires a formal description of the laser induced ultrafast magnetization dynamics which involves the multiple
possible interactions between the photons, electrons, the spins and the lattice together with the complexity of the electronic band structures of the magnetic materials.

Several models have been proposed to describe the demagnetization process occurring in the sub-picosecond time scale. The
first one  involves three interacting baths at different temperatures corresponding to the charges, the spins and the lattice which are out of equilibrium \cite{JYB2001}. A quantum model, including the effects of exchange and spin-orbit interactions, has accounted for spin flips in terms of their dephasing after they redistribute in the excited states \cite{Hubner1998,Zhang2002,zhang2008}. An extension of this model including the combined interaction of the laser and spin-orbit successfully described elementary spin-flips or so-called lambda processes \cite{zhang2000}. Modeling the spins dynamics at later times,
a few picoseconds after laser excitation, has reached a consensus. In that case, the spin-phonon interaction prevails and is responsible for the
damping of the precession of the magnetization in ferromagnets exception made when approaching the Curie temperature for which long range
fluctuations maintain a non-equilibrium spin bath for a long time \cite{chubykalo-fesenko2006,Oppeneer2004}. A revival of the debate regarding the origin of the ultrafast demagnetization occurred with the assumption that spin flips could be due to a
mechanism similar to the Elliott-Yafet scattering of conduction electrons by magnetic impurities
\cite{koopmans2005Unifying,koopmans2007Ultimate}. Some experiments do not support this model for example regarding the effect of
magnetic impurities on the spin dynamics in doped ferromagnetic transition metals \cite{Radu-Back 2005}.

More recently a new controversy started regarding the interpretation of the time resolved magneto-optical response as a signature of the magnetization dynamics \cite{Oppeneer2011,Zhang2011}. The main origin of this controversy lies in the distinctions that one ought to make between the time dependent response function (magneto-optical signal) and the system's dynamics (the spins populations). In the present work we show that there is a straightforward manner to clarify the debate by considering separately the coherent and population dynamics in the magneto-optical response. Indeed it is known that in metals coherent magnetism is important as shown in a recent detailed study of the charges and spins dynamics performed in Ni and CoPt$_{3}$ ferromagnetic films \cite{Bigot2009}.

Let us consider the simplest possible system constituted of eight quantum levels interacting with the laser field such that the interaction takes into account the relativistic corrections to the quantum electron dynamics, including the spin-orbit interaction. We determine the response function from the density matrix formalism including the time ordered third order nonlinear terms. This approach allows understanding the main differences between the coherences and spins populations.
We consider a one-electron Hamiltonian with an effective Coulomb interaction perturbed by the spin-orbit and kinetic momentum-laser vector potential interactions. Using such Hamiltonian, applied to the band structure of metals, P. N. Argyres has shown that static magneto-optical Kerr and Faraday effects in
ferromagnetic materials mainly come from spin-orbit interaction with the ionic field \cite{argyres1955}. For the purpose of understanding the main steps of our approach, which primarily aims at determining the respective roles played by the coherent versus population dynamics in the response function (Faraday rotation), we consider the case of a simple discrete eight level system representing a reduced hydrogen-like system.

\noindent The relativistic contributions to the ultra-fast magneto-optical dynamics are considered via the Foldy-Wouthuysen
transformation of the Dirac equation which enlightens the various interaction terms between spins and femtosecond laser fields. We add to Argyres' approach the terms of the Foldy-Wouthuysen transformation to second order in $\frac{1}{m}$ for the electron from an Hydrogen-like atom submitted to different static fields : the ionic field $\textbf{E}_{i}$ and its associated central ionic potential
$V_{i}(\textbf{r})$ and a strong static homogeneous magnetic field $B_{M}\textbf{e}_{z}$ with potential vector $\textbf{A}_{M}=-\frac{1}{2}\textbf{R}
\wedge \textbf{B}_{M}$. The electron is interacting with a laser field described in the Coulomb gauge associated to an homogeneous electric field
$\textbf{E}_{L}$, to its colinear vector potential $\textbf{A}_{L}$ and related magnetic field $\textbf{B}_{L}$.

\noindent The Hamiltonian $H_{0}$ corresponding to no interaction with the laser reflects the Zeeman splitting in a strong magnetic field. At this point the degeneracy of the spin states are already lifted and the effect of spin-orbit interaction is to slightly shift some energy levels.
\begin{eqnarray}
H_{0}&=&\frac{1}{2m}(\textbf{p}-q\textbf{A}_{M})^{2}+q V_{i}(\textbf{r})-\frac{q \hbar^{2}}{8 m^{2} c^{2}} \nabla \cdot \textbf{E}_{i}\\\nonumber
&-&\frac{q}{m}\textbf{S}\cdot \textbf{B}_{M}-\frac{q}{2 m^{2} c^{2}} \textbf{S} \cdot [\textbf{E}_{i} \wedge (\textbf{p}-q\textbf{A}_{M})]\nonumber
\end{eqnarray}

\noindent Neglecting the second order terms in the laser vector potential the interaction Hamiltonian is :

\begin{eqnarray}
H_{int}&=&-\frac{q}{m}\textbf{$\Pi$}\cdot \textbf{A}_{L}-\frac{q}{2m^{2}c^{2}}[(\textbf{p}-q\textbf{A}_{M})\wedge\textbf{S}]\cdot
\textbf{E}_{L}\\\nonumber &-&\frac{q}{m}\textbf{S}\cdot \textbf{B}_{L}-\frac{\imath q \hbar}{4 m^{2} c^{2}}\textbf{S}\cdot(\nabla
\wedge\textbf{E}_{L})\nonumber
\label{hamiltonien_interaction}
\end{eqnarray}
\noindent With the kinetic momentum operator \mbox{\boldmath$\Pi$$=\frac{m}{\imath \hbar}[\textbf{R},H_{0}]$} :

\begin{eqnarray}
\mbox{\boldmath$\Pi$}=\textbf{p}-q\textbf{A}_{M}+ \frac{q}{2 m c^{2}}\textbf{S}\wedge\textbf{E}_{i}
\end{eqnarray}

\begin{figure}[!h]
    \includegraphics[scale=0.3]{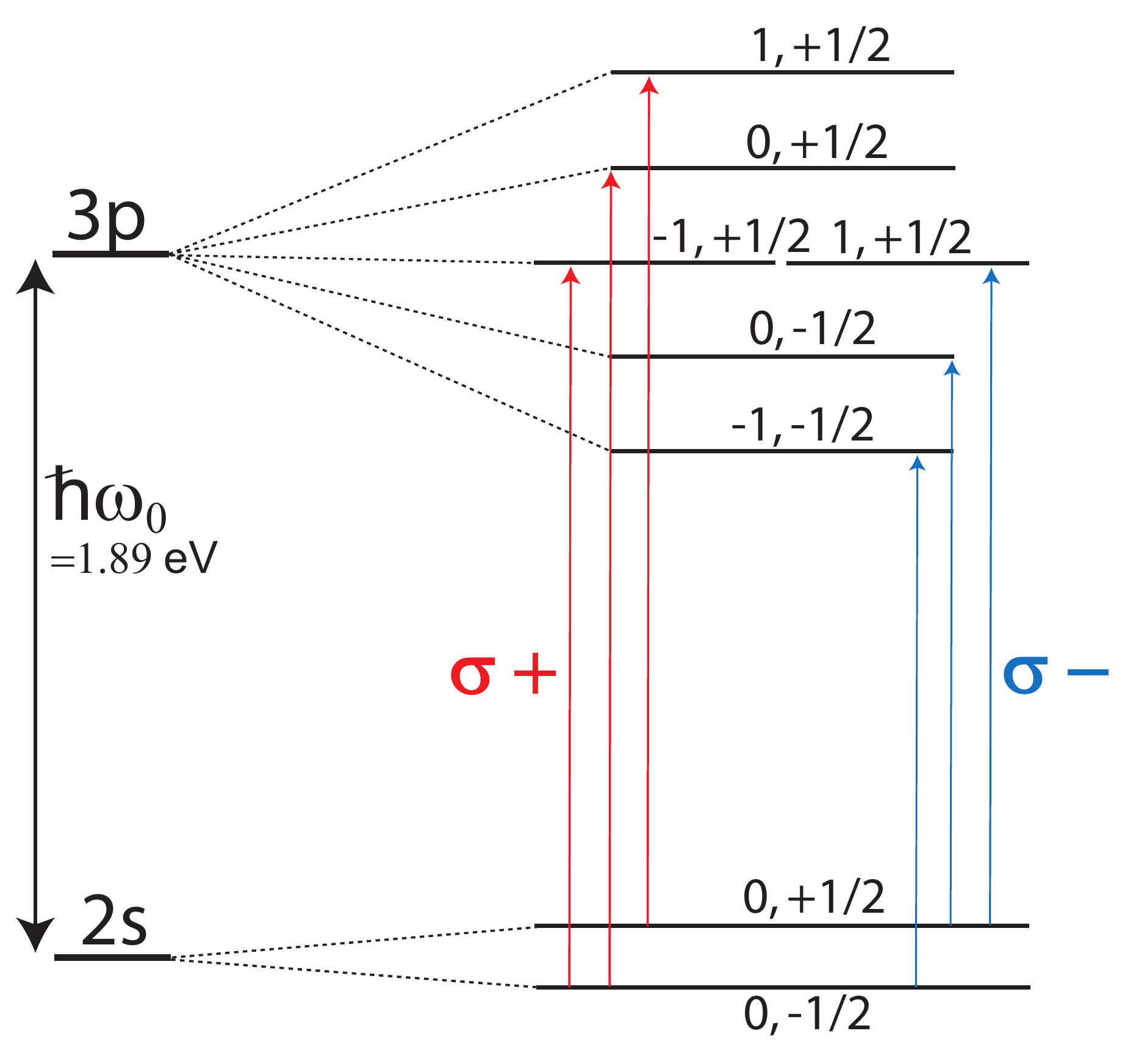}
    \caption{\label{niveaux} Considered transitions in the Hydrogen-like atom. For each level we indicate $l_{z},s_{z}$. $\sigma_{\pm}$ stands for a circularly polarized field along $\textbf{e}_{x} \pm \imath \textbf{e}_{y}$.}
\end{figure}

\noindent
We apply the electric dipolar approximation to the interaction Hamiltonian and consider the $2s$ and $3p$ levels of the Hydrogen atom ; their energy difference without spin-orbit coupling and static magnetic field is associated to the frequency $\omega_{0}$ as shown in figure \ref{niveaux}. This energy is close to the typical photon energies used in experiments. We do not consider the magnetic dipolar interaction term which is off-resonant. The two first terms of equation 2 imply a variation of the projection of the orbital momentum along the quantization axes $\Delta l=\pm 1, \Delta l_{z}=\pm 1, \Delta s_{z}=0 $ and the terms containing the spin ($H_{\alpha}$ and $H_{\beta}$) and spin-flip transitions with $\Delta l=\pm 1, \Delta l_{z}=0, \Delta s_{z}=\pm 1 $. Table \ref{comparaison} shows the orders of magnitude of each interaction matrix element compared to the predominant interaction term \mbox{$<j|\frac{q}{m}\textbf{p}.\textbf{A}_{L}|i>$} for two states $|i>$ and $<j|$ among the eight states sketched at figure 1. In the case of our hydrogen-like model, we will see that the contributions  \mbox{$H_{\alpha}=\frac{q^{2}}{2 m^{2} c^{2}}[\textbf{S}\wedge \textbf{E}_{i}].\textbf{A}_{L}$} and \mbox{$H_{\beta}=\frac{q}{2 m^{2} c^{2}}[\textbf{p}\wedge \textbf{S}].\textbf{E}_{L}$} can be neglected in the magneto-optical response but not in the spin dynamics. It is not the purpose to discuss here their influence in the case of a ferromagnetic solid, where additional phenomena such as a dynamical anisotropy can be induced by the presence of an external field \cite{Stoehr2009}, the laser field for example in our case.

\begin{table}
\caption{\label{comparaison} Scale of the considered interaction matrix elements in Hydrogen compared to the canonical momentum-laser vector potential interaction matrix element.}
\begin{ruledtabular}
\begin{tabular}{rlll}
&$<j|\frac{q}{m}p.A_{L}|i>$ & $\simeq$&$1$ \\
&$<j|\frac{q^{2}}{m}A_{M}.A_{L}|i> $ &$\simeq$&$ 3.10^{-5}B_{M}$ \\
$<j|H_{\alpha}|i>=$&$<j|\frac{q^{2}}{2 m^{2} c^{2}}S\wedge E_{i}.A_{L}|i>$ & $\simeq$&$ 1.10^{-6}$ \\
$<j|H_{\beta}|i>=$&$<j|\frac{q}{2 m^{2} c^{2}}p\wedge S.E_{L}|i> $& $\simeq$&$ 9.10^{-7}$ \\
\end{tabular}
\end{ruledtabular}
\end{table}

 In order to model magneto-optical pump-probe experiments, the evolution of the system is calculated in the Liouville formalism to the third order of the laser perturbation \cite{Mukamel}. The relaxation time of the coherences ($\rho_{nm},n\neq m$) is $T_{2}$ whereas it is $T_{1}$ for the population differences ($\rho_{nn}-\rho_{mm}, n \in {3,8}; m=1,2$). For simplicity they are assumed to be the same for each transitions. The electric field of the laser pulses is given by \mbox{$\textbf{E}_{L}(t)=\frac{1}{2}[\epsilon(t,\tau) e^{-\imath\omega_{L}t}+\epsilon^{*}(t,\tau)e^{\imath\omega_{L}t}]\textbf{e}_{x}$} where $\epsilon(t,\tau)$ is a gaussian centered on the pump-probe delay $\tau$ or 0 in the case of the probe or pump fields. The laser frequency $\omega_{L}$ is equal or close to $\omega_{0}$. The effective dipolar moment of the system $\textbf{D}$ gives the first and third order polarizations of the atom $\textbf{P}^{(1)}(t)=Tr(\rho^{(1)} \textbf{D})$ and $\textbf{P}^{(3)}(t,\tau)=Tr(\rho^{(3)} \textbf{D})$. The polarization is calculated using the rotating wave approximation and depends on both the time $t$ and delay $\tau$ between the pump and probe pulses. As we do not consider any propagation effect in this simple atomistic approach, we set arbitrarily that the radiated electric field at order \textit{(n)} is $\textbf{E}^{(n)}\equiv \textbf{P}^{(n)}$.
 We consider pump and probe linearly polarized in the plane perpendicular to the quantization axes and calculate the dynamical magneto-optical rotation induced by the sample in the Jones formalism. In order to follow the common experimental procedure for magneto-optical measurements with a polarization bridge, we define $\textbf{E}^{\prime(3)}(t,\tau)$ as the rotated $\textbf{E}^{(3)}(t,\tau)$ due to a half-wave plate tilted by an angle of $\frac{\pi}{8}$ with respect to the $(\textbf{e}_{x}$, $\textbf{e}_{y})$ axes, the measured differential intensities $\Delta I_{x, (y)}$ are:

\begin{eqnarray}
\label{intensite}
\Delta I_{x,(y)}=\int_{-\infty}^{+\infty}2\Re[\textbf{E}^{\prime(3)}_{x,(y)}(t,\tau)\cdot\textbf{E}^{\prime*(1)}_{x,(y)}(t)]dt
\end{eqnarray}

 The differential rotation for a positive magnetic field $\Delta\Theta_{+B_{M}}(\tau)=[\Delta I_{x}-\Delta I_{y}]_{+B_{M}}$ is linked to the first and third order field's amplitude and to their instantaneous rotations $\Theta_{+B_{M}}^{(1)}(t)$ and $\Theta_{+B_{M}}^{(3)}(t,\tau)$; here we assume the ellipticity to be negligible.

\begin{eqnarray}
\label{rotation}
\Delta\Theta_{+B_{M}}(\tau)=&\int_{-\infty}^{+\infty} sin[\Theta_{+B_{M}}^{(3)}(t,\tau)+\Theta_{+B_{M}}^{(1)}(t)]\\\nonumber
&\times 2|\textbf{E}^{(3)}_{x,(y)}(t,\tau)||\textbf{E}^{*(1)}_{x,(y)}(t)|]dt\nonumber
\end{eqnarray}

The resulting magneto-optical signal is then proportional to: $\Delta\Theta(\tau)=\Delta\Theta_{+B_{M}}(\tau)-\Delta\Theta_{-B_{M}}(\tau)$ obtained for the two directions $\pm{B_{M}}$ of the magnetic field. As done experimentally, we perform a differentiation on the magnetic field and normalize the resulting rotation with the linear magneto-optical rotation $\Theta_{\pm B_{M}}$:

\begin{eqnarray}
\frac{\Delta\Theta(\tau)}{\Theta}=\frac{\Delta\Theta_{+B_{M}}(\tau)-\Delta\Theta_{-B_{M}}(\tau)}{\Theta_{+B_{M}}-\Theta_{-B_{M}}}
\label{rotation_mesuree}
\end{eqnarray}

\begin{figure}
      \includegraphics[scale=0.7]{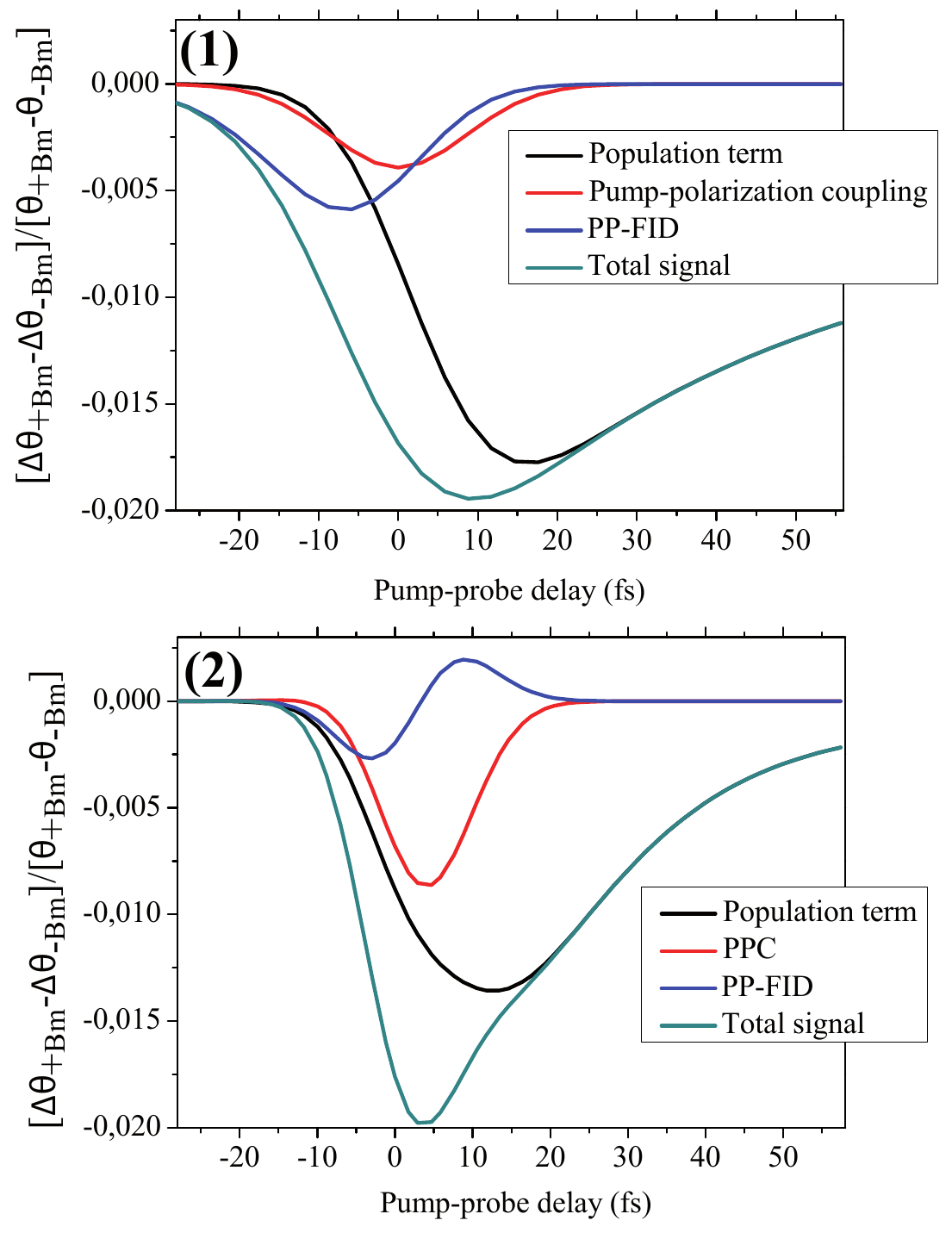}
       \caption{\label{termes_pompe_sonde}  Magneto-optical rotation for the population (black), pump-polarization-coupling (red), pump-perturbed free-induction decay (blue) and total (green) signals. (1): on resonance ($\hbar\omega_{L}=\hbar\omega_{0}=1.89 eV$) and (2): off resonance ($\hbar\omega_{L}=  1.51 eV$).}
\end{figure}

Considering the time ordering of a single probe and two pump pulses one can distinguish three terms corresponding to different coupling between pump, probe and the polarization of the system as described in the case of charge dynamics by Brito-Cruz et al. \cite{Brito-Cruz_1988}. First, the time ordering of the fields  \mbox{$\epsilon_{pump}(t=0)\epsilon_{pump}^{*}(t=0)\epsilon_{probe}(t=\tau)$} describes the "population dynamics" induced by the pump field $\epsilon_{pump},\epsilon_{pump}^{*}$. It is maximal for positive pump-probe delays and relaxes with the population decay time $T_{1}$. The second term named "pump-polarization coupling" (PPC) is given by the sequence \mbox{$\epsilon_{pump}^{*}(t=0)\epsilon_{probe}(t=\tau)\epsilon_{pump}(t=0)$} which corresponds to a convolution of the pump pulse with the exponential decay of the density matrix's coherences within the $T_{2}$ time.
The third term \mbox{$\epsilon_{probe}(t=\tau)\epsilon_{pump}^{*}(t=0)\epsilon_{pump}(t=0)$} is maximum at negative pump-probe delays and corresponds to the coherences generated by the probe which couple to the pump, sometimes also named "pump-perturbed free induction decay" (PP-FID).
These two latter terms, also named "coherent terms" hereafter, are strongly related to the dephasing of the polarization as pump and probe can only couple if the density matrix's coherences are non-zero. Figure \ref{termes_pompe_sonde} shows the dynamical differential magneto-optical rotation calculated for each of these three terms using the interaction Hamiltonian in equation 2.

In the numerical simulations, we have considered gaussian pulses of width 10 fs and a probe field 10 times less intense than the one of the pump. The dephasing time of the coherences  and the lifetime of the differences of population are chosen to be 10 fs and 100 fs. The static magnetic field is 1 Tesla. The pump and probe laser fields are taken linearly polarized along $\textbf{e}_{x}$. The unperturbed ground-level populations are 0.9 for $s_{z}=\mp\frac{1}{2}$ and 0.1 for $s_{z}=\pm\frac{1}{2}$ for a magnetic field $\pm B_{M}$. These results clearly show that the coherent terms are important in the magneto-optical signals and that their contributions respective to the spins population dynamics depends on the laser detuning.

The next issue that we address now is the difference between the pump-probe magneto-optical signal and the spin and orbital momentum dynamics. Towards that purpose we determine two different quantities: the projection of the spin and the orbital momentum operators $\langle{S_{z}^{(2)}}\rangle$ and $\langle{L_{z}^{(2)}}\rangle$ along the quantification axis $\textbf{e}_{z}$. Note that the dynamics is now represented by the trace of the operators multiplied by the second order nonlinear terms of the density matrix as they are related to the density matrix's populations. The spin dynamics are due to the $H_{\alpha}$ and $H_{\beta}$ spin-flip terms. Figure 3 shows their dynamics for the three time orderings of the pump and probe fields. The population terms (fig.3a and 3b) are represented as a function of time $\textit{t}$ as they do not depend on the probe field. The two coherent terms are integrated over $\textit{t}$ and represented as a function of pump-probe delay $\tau$, fig.3c and 3d for  "PPC" and fig.3e and 3f for "PP-FID" terms, as they explicitly depend on time ordered sequences involving both the pump and probe pulses. $\langle{S_{z}^{(2)}}\rangle$ and $\langle{L_{z}^{(2)}}\rangle$ are respectively real and imaginary because they are generated by the pulses $\epsilon_{pump}^{*}$ and $\epsilon_{probe}$. In order to calculate real quantities one should also have considered $\epsilon_{pump}$ and $\epsilon_{probe}^{*}$ which does not correspond to the experimental configuration chosen here.
\begin{figure}
\begin{tabular}{cc}
    \includegraphics[scale=0.4]{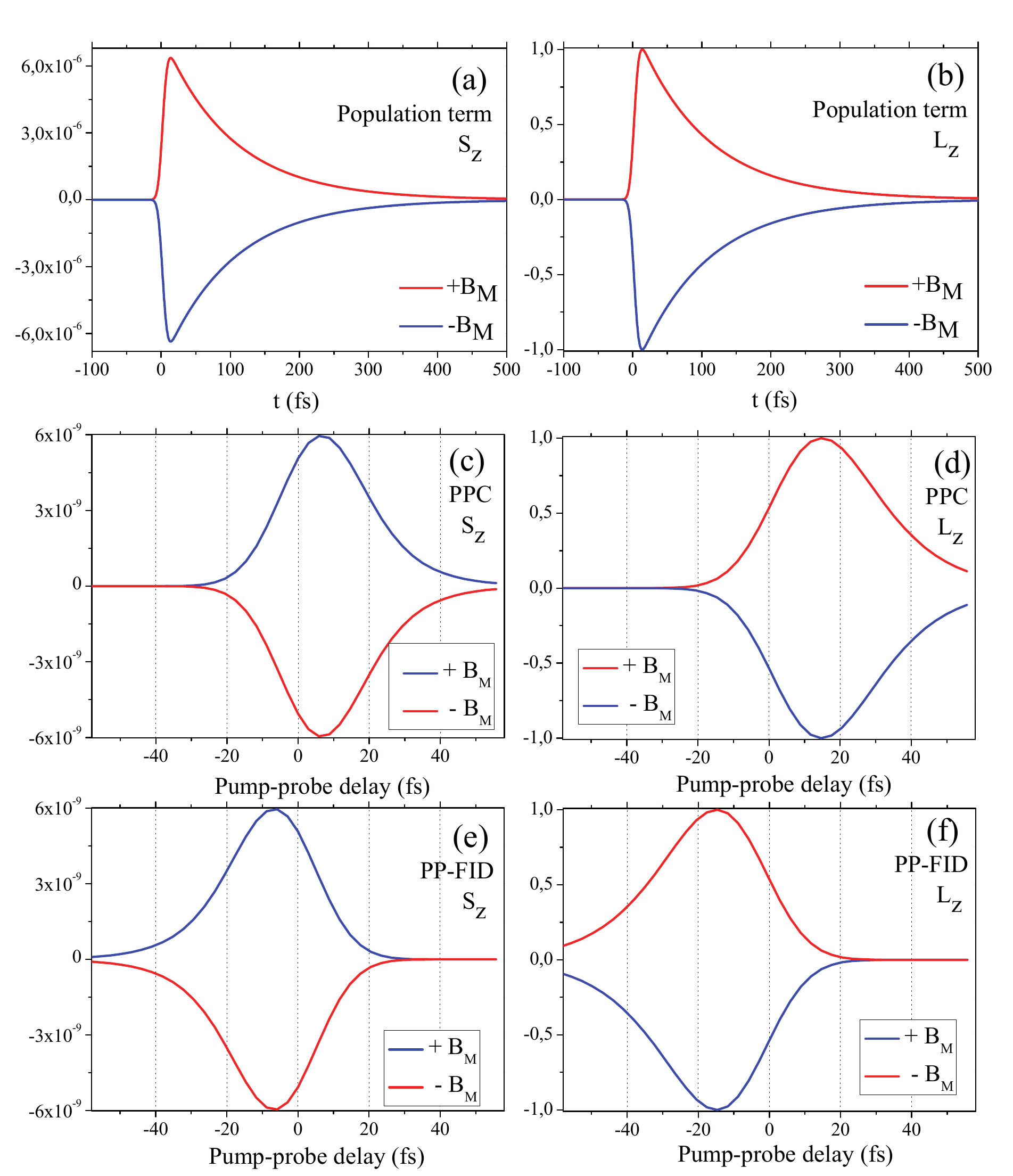}
\end{tabular}
\caption{\label{magnetization}  Spin and orbital momentum dynamics for the three pump-probe field sequences. (a): $Tr[\rho^{(2)} S_{z}(t)]$ and (b): $Tr[\rho^{(2)} L_{z}(t)]$ for the population term. (c): $\Re\{\int dt Tr[\rho^{(2)} S_{z}(t,\tau)]\}$ and (d): $\Im\{\int dtTr[\rho^{(2)} L_{z}(t,\tau)]\}$ for the pump-polarization coupling term. (e) and (f): same quantities as (c) and (d) but for the pump-perturbed free-induction decay term. (c) to (f) have the same normalization factor.}
\end{figure}

In conclusion, we have shown that in an ultra-fast magneto-optical experiment the time ordering of the pulses has to be taken into account especially in order to distinguish the coherent response from the populations dynamics. Out of resonance the magneto-optical coherent signal increases. More importantly the spin and orbital momentum's dynamics, due to the second order in perturbation of the density matrix, both participate to the coherent magneto-optical response. Extrapolating the present results to more complex magnetic systems shows that the coherent spin-photon interaction can be used to manipulate the magnetization of spin devices at the femtosecond time scale.

 The authors thank Y. Hinschberger, P.-A. Hervieux and G. Lefkidis for fruitful discussions. H. V. is grateful to the computing department of the IPCMS and J.-Y. B. acknowledges the financial support of the European Research Council with the ERC Advanced Grant ATOMAG (ERC-2009-AdG-20090325 247452).

\end{document}